\documentclass[prb, twocolumn, floatfix]{revtex4}
\usepackage{hyperref}
\usepackage{amssymb}
\usepackage{amsmath}
\usepackage{float}
\usepackage{bbm}
\usepackage{graphicx}
\usepackage{epsfig}
\usepackage{epstopdf}

\allowdisplaybreaks[0]

\newcommand{\II}{I\otimes I}
\newcommand{\IX}{I\otimes\sigma_x}

\newcommand{\XI}{\sigma_x\otimes I}

\newcommand{\XZ}{\sigma_x\otimes\sigma_z}

\newcommand{\ZX}{\sigma_z\otimes\sigma_x}

\newcommand{\ZZ}{\sigma_z\otimes\sigma_z}
\newcommand{\nn}{\nonumber}

\begin{document}

\title{Improving the gate fidelity of capacitively coupled spin qubits}

\author{Xin Wang$^{1}$, Edwin Barnes$^{1,2}$, and S.~Das Sarma$^{1,2}$}

\affiliation{$^{1}$Condensed Matter Theory Center, Department of Physics,
University of Maryland, College Park, Maryland 20742, USA\\
$^{2}$Joint Quantum Institute, University of Maryland, College Park, Maryland 20742, USA}

\date{\today}

\begin{abstract}

Capacitively coupled semiconductor spin qubits hold promise as the building blocks of a scalable quantum computing architecture with long-range coupling between distant qubits. However, the two-qubit gate fidelities achieved in experiments to date have been severely limited by decoherence originating from charge noise and hyperfine interactions with nuclear spins, and are  currently unacceptably low for any conceivable multi-qubit gate operations.
Here, we present control protocols that implement two-qubit entangling gates while substantially suppressing errors due to both types of noise. These protocols are obtained by making simple modifications to control sequences already used in the laboratory and should thus be easy enough for immediate experimental realization.
Together with existing control protocols for robust single-qubit gates, our results constitute an important step toward scalable quantum computation using spin qubits in semiconductor platforms.

\end{abstract}

\maketitle

\section{Introduction}

Precise manipulation of coupled quantum systems is at the heart of quantum technologies. Quantum gates acting on two or more qubits are the building blocks of  quantum algorithms and  the engines of entanglement creation, the key feature which sets apart many technological applications such as quantum information processing and quantum communication\cite{NielsenChuang.00} from their classical counterparts. The ability to manipulate two or more interacting qubits has been demonstrated with reasonably high fidelity in many systems such as superconducting qubits,\cite{DiCarlo.09} trapped ions,\cite{Gulde.03} and optical systems,\cite{O'Brien.03}
but for the purpose of quantum computing, scaling to many qubits may be ultimately easier to achieve with semiconductor spin qubits\cite{Nowack.11,Veldhorst.14,Kim.14,Muhonen.14} because they are nanoscale devices that can potentially take advantage of the preexisting infrastructure for fabricating semiconductor microchips.\cite{Taylor.05} The singlet-triplet qubit,\cite{Petta.05} which encodes a qubit in two-electron spin states, bears the advantage that it is immune to homogeneous fluctuations in the external magnetic field and is the simplest spin qubit that can be controlled purely electrically. While it has had vast success at the single-qubit level with long coherence times and high control fidelities,\cite{Foletti.09,Bluhm.11,Shulman.14} the implementation of two-qubit gates has been challenging and is currently the bottleneck preventing progress towards demonstrations of quantum algorithms, which necessitate both single- and multi-qubit gate operations.

Attempts to design high-fidelity two-qubit gates for singlet-triplet qubits have mostly assumed that the qubits are coupled via the exchange interaction,\cite{Klinovaja.12,Kestner.13} in which case one may focus on one pair of exchange-coupled spins at a time, a great simplification from the original four-spin problem. Despite its theoretical simplicity and elegance, this approach is hard to implement in experiments with multiple gate-defined quantum dots due to difficulties in addressing a single exchange coupling in the array.  An alternative proposal couples two singlet-triplet qubits with a capacitive interaction, i.e. one may alter the electrostatic potential, and hence the precession rate, in the ``target" qubit by changing the electron charge configuration of the ``control'' qubit.\cite{Taylor.05,vanWeperen.11} This proposal allows for long-range, individually controllable couplings between qubits,\cite{Trifunovic.12} which is important for multi-qubit devices, but also has a serious difficulty in that the capacitive coupling is typically much weaker than the exchange coupling and is much more susceptible to noise. An experimental breakthrough in this problem came from the employment of a ``simultaneous dynamical decoupling'' sequence which cancels part of the noise-induced evolution error, leading to the first demonstration of a singlet-triplet two-qubit gate, with a fidelity of 72\%.\cite{Shulman.12} In order to meet the requirement for fault-tolerant quantum computation, one must further reduce the error either at the hardware level by enhancing the capacitive coupling or making better noise-free samples, or at the software level using dynamically corrected gates. Several theoretical works have employed the latter approach to design higher-fidelity single-qubit gates.\cite{Wang.12,Grace.12,Kestner.13,Wang.14a,Barnes.14}

Despite the experimental advance demonstrated in Ref.~\onlinecite{Shulman.12}, theoretical research on improving the fidelity of singlet-triplet two-qubit gates via capacitive coupling has remained rare. While the inter-qubit coupling term has been individually treated in NMR{\cite{Jones.03}}, the unavoidable presence of single-qubit terms and their associated errors complicates the problem considerably, making previous approaches inapplicable. Further difficulty arises that capacitive coupling exhibits less symmetry compared with the exchange coupling, preventing one from factorizing the larger two-qubit Hilbert space into separate subspaces. This makes it difficult to adapt (or generalize) dynamically corrected gates developed for single-qubit operations\cite{Wang.12,Grace.12,Kestner.13,Wang.14a,Barnes.14} to design a robust two-qubit gate for capacitive coupling as was done in the case of exchange coupling.\cite{Kestner.13,Wang.14a} Still, it has recently been realized that one may systematically generate entangling gates with capacitive coupling which are equivalent to the well-known CNOT gate by using a single square pulse.\cite{Calderon.14}  In addition, the decoherence mechanisms relevant for the protocol implemented in Ref.~\onlinecite{Shulman.12} have been further investigated in Ref.~\onlinecite{Srinivasa.14}. However, a systematic way of performing two-qubit gate operations that are robust against noise is still an important open question. Here, we address this issue by showing that minor, precise modifications to the sequence already experimentally implemented in Ref.~\onlinecite{Shulman.12} can further cancel the effects of noise and significantly enhance the two-qubit gate fidelity. We start with the simpler case in which the magnetic field gradient is much smaller than the control field; using the results we obtain in this limit as a guide, we then solve the problem for more general parameter regimes. We show how to systematically generate pulse sequences that implement entangling two-qubit gates while suppressing both nuclear and charge noise errors by more than an order of magnitude.

\section{Results}
\subsection{Model}

We consider two singlet-triplet qubits (labelled by $A$ and $B$) coupled capacitively. The Hamiltonian reads\cite{Shulman.12}
\begin{align}
H(\{J^A,&h^A\},\{J^B,h^B\}, J^{AB})=(J^A\sigma_z+h^A\sigma_x)\otimes I\nn\\
&+I\otimes(J^B\sigma_z+h^B\sigma_x)+J^{AB}\ZZ.\label{eq:twoqHam}
\end{align}
Here, $J^A$ ($J^B$) is the exchange interaction between the two spins on qubit $A$ ($B$) and determines the rate of single-qubit $z$-rotations. $h^A$ ($h^B$) is the magnetic field gradient enabling single-qubit $x$-rotations. $J^{AB}$ denotes the capacitive coupling between the two qubits which creates entanglement, and is typically much smaller than $J^A$ and $J^B$. In practice, the exchange interactions are controlled by the bias voltages (detunings) applied to each qubit, $J^A=J^A(\epsilon^A)$ and $J^B=J^B(\epsilon^B)$. One can similarly write $J^{AB}=J^{AB}(\epsilon^{AB})$ where $\epsilon^{AB}$ may be defined as a linear combination of $\epsilon^A$ and $\epsilon^B$. The evolution under the Hamiltonian, Eq.~\eqref{eq:twoqHam}, for time $t$ is thus a two-qubit gate denoted as:
\begin{align}
U(\{J^A,h^A\},\{J^B,h^B\}, J^{AB},t)\notag\\=e^{-iH(\{J^A,h^A\},\{J^B,h^B\}, J^{AB})t}.\label{eq:onepieceevol}
\end{align}

Similarly to the single-qubit case, the evolution of two qubits is subject to two noise channels. In the case where the qubits are in semiconductors like GaAs,\cite{Johnson.05,Bluhm.11,Medford.12} the nuclear noise arises from the flip-flops of surrounding nuclear spins mediated by the hyperfine interaction with the electron spins in the quantum dots, which changes $h^A$ to $h^A+\delta h^A$ and $h^B$ to $h^B+\delta h^B$ (Ref.~\onlinecite{Reilly.08}). On the other hand, charged impurities near the confined electrons lead to fluctuations of the confinement potential and consequently the exchange interaction. The shift in the exchange interaction can be expressed as $J^A\rightarrow J^A+g^A(J^A)\delta\epsilon^A$, $J^B\rightarrow J^B+g^B(J^B)\delta\epsilon^B$ and $J^{AB}\rightarrow J^{AB}+g^{AB}(J^{AB})\delta\epsilon^{AB}$, where $g$ is determined by the dependence of $J$ on the detuning $\epsilon$ (Ref.~\onlinecite{Dial.13}). Together these errors give rise to inaccuracies in the resulting two-qubit quantum gates, namely $U\rightarrow U+\delta U$ with $\delta U$ dependent on the $\delta h$'s and $\delta\epsilon$'s. The goal of this paper is to reduce the effect of noise (i.e. $\delta U$) using optimized pulse sequences. In this work we also make the static noise approximation, namely the $\delta h$'s and $\delta\epsilon$'s are assumed to be unknown constants for a given run of a quantum gate, but may change from one run to the next. The static noise approximation is valid in most experimental spin qubit situations since the electrically controlled gate operations can typically be very fast compared with the slow noise fluctuations in the environment.\cite{Medford.12,Dial.13} For spin qubits in silicon, the nuclear noise is absent and one has only charge noise to deal with.\cite{Maune.12,Wu.14,Wang.14b} Our results are applicable in this case as well.

The explicit analytical form of Eq.~\eqref{eq:onepieceevol} for arbitrary parameters is complicated and involves solutions to a quartic eigenequation (see Methods), making it difficult to find pulse sequences which implement well-known entangling gates such as the CNOT. In Ref.~\onlinecite{Shulman.12}, a two-qubit gate was experimentally demonstrated by employing a ``simultaneous dynamical decoupling'' sequence given by
\begin{align}
U&(\{J^A,0\},\{J^B,0\}, J^{AB},t)e^{-i\frac{\pi}{2}(\IX+\XI)}\nn\\
&\times U(\{J^A,0\},\{J^B,0\}, J^{AB},t)\notag\\&=
\begin{pmatrix}
0 & 0 & 0 & -e^{-i\phi} \\
0 & 0 &  -e^{i\phi}  & 0 \\
0 &  -e^{i\phi}  & 0 & 0 \\
 -e^{-i\phi}  & 0 & 0 & 0 \\
\end{pmatrix},\label{eq:one-piece}
\end{align}
where $\phi=2J^{AB}t$, and one makes the approximation that $h^A$ and $h^B$ are much weaker compared to the control fields and are thus negligible when those fields are pulsed on. The resulting gate has a particularly useful anti-diagonal form. One typically uses the Makhlin invariants\cite{Makhlin.02} to classify two-qubit gates such that two gates with identical invariants may be converted into each other using only single-qubit (local) operations. The Makhlin invariants for Eq.~\eqref{eq:one-piece} are $G_1=\cos^22\phi$ and $G_2=2+\cos4\phi$. For $\phi=N\pi/4$ with $N$ an odd integer, it is equivalent to CNOT ($G_1=0, G_2=1$) up to local operations.

\begin{figure*}
\centering\includegraphics[width=1.8\columnwidth]{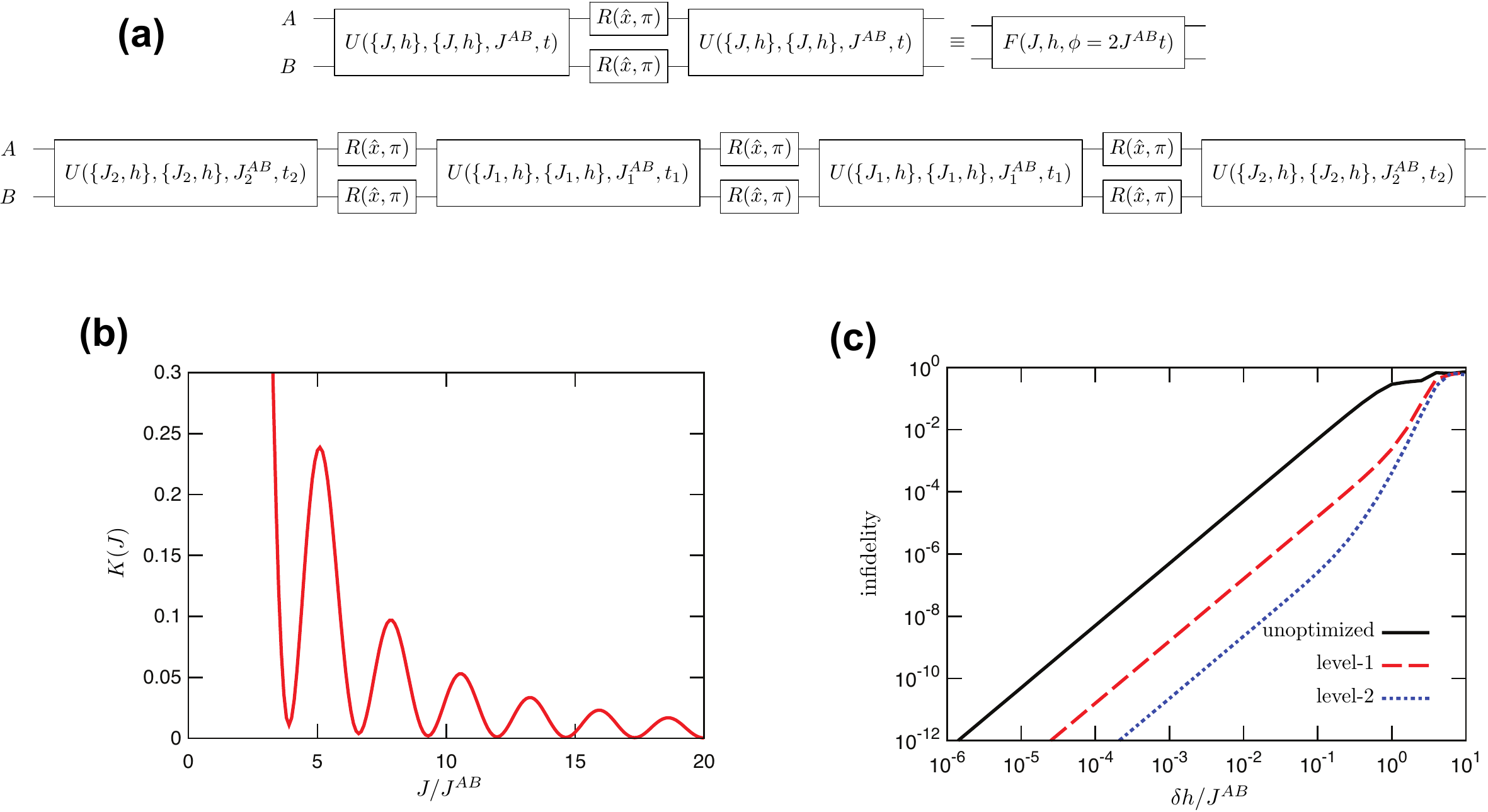}
\caption{Nuclear noise reduction for $h=0$. The charge noise has been set to zero. (a) Quantum circuits showing the level-1 and level-2 pulse sequences used to reduce the nuclear noise. (b) The cost function, Eq.~\eqref{eq:Konepiece}. (c) Infidelity v.s. nuclear noise $\delta h/J^{AB}$ for unoptimized (black line), level-1 optimized (red dashed line) and level-2 optimized (blue dotted line) sequences. Parameters:
Unoptimized pulse: $J=3J^{AB}$, $\phi=3\pi/4$ $[t=3\pi/(8J^{AB})]$;
Optimized 1-piece pulse: $J=9.2901J^{AB}$, $\phi=3\pi/4$ $[t=3\pi/(8J^{AB})]$;
Optimized 2-piece pulse: $J_1=18.767J^{AB}$, $J_2=9.2123J^{AB}$, $t_1=0.33606/J^{AB}$, $t_2=3\pi/(8J^{AB})-t_1$.}
\label{fig:dhonly}
\end{figure*}

In this paper, we start from Eq.~\eqref{eq:one-piece} and show how one can reduce the gate error due to nuclear noise and charge noise. For the first half of our results, we make the same assumption $h^A, h^B\ll J^A,J^B$,$J^{AB}$ that has been utilized and verified in experiments.\cite{Shulman.12} In this relatively simple case, we show that one may construct dynamically corrected two-qubit gates in an intuitive way. Later on, we discuss the more complicated yet practical case with nonzero $h^A$ and $h^B$, where it is harder to gain intuition even in the absence of noise, and extra work is needed to ensure the equivalence between the outcome of the pulse sequence and the desired two-qubit gate (CNOT) by direct evaluation of the Makhlin invariants. Nevertheless, we explicitly demonstrate that pulse sequences can be constructed in this demanding situation as well using constrained numerical optimization techniques that start from the solution to the simpler $h^A=h^B=0$ case.

We also make a few nonessential assumptions for the purpose of presentation. We show results for ``symmetric'' pulsing, namely $J^A=J^B=J$ and $h^A=h^B=h$.
Regarding the charge noise, we assume the empirical form\cite{Dial.13} $J^A=J^A_0\exp(\epsilon^A/\epsilon_0)$ so that $\delta J^A=J^A\delta\epsilon^A/\epsilon_0$, and similarly $\delta J^B=J^B\delta\epsilon^B/\epsilon_0$. For the capacitive coupling $J^{AB}$, the mechanism of charge noise is different than for individual qubits. While in Ref.~{\onlinecite{Shulman.12}} it has been argued empirically that $J^{AB}\propto J^AJ^B$, no microscopic calculation of this relationship has appeared in the literature, and the precise form is likely sample-dependent.  We therefore assume the simple form $\delta J^{AB}=J^{AB}\delta\epsilon^{AB}/\epsilon_0$.
While we make these assumptions because concrete parameter values are needed in evaluating the performance of the gates, we emphasise that the applicability of our method does not depend in any way on these assumptions. In fact, all these restrictions can be lifted either by modifying the cost function or altering the parameters used in the numerical search. For example, deviations from the assumed exponential model of charge noise may easily be accommodated according to the method presented in Ref.~{\onlinecite{Wang.14a}}.

\subsection{Nuclear noise reduction for $h=0$ case}

We write the evolution operator, Eq.~\eqref{eq:one-piece}, under nuclear noise $\delta h^A$ and $\delta h^B$ as
\begin{align}
&\widetilde{F}(J, h,\phi,\delta h^A,\delta h^B)\nn\\
=&\ U(\{J,h+\delta h^A\},\{J,h+\delta h^B\}, J^{AB},\phi/J^{AB})\nn\\
&\times e^{-i\frac{\pi}{2}(\IX+\XI)}\notag\\
&\times U(\{J,h+\delta h^A\},\{J,h+\delta h^B\}, J^{AB},\phi/J^{AB}).
\label{eq:one-piecewdh}
\end{align}
We also define the noise-free component of Eq.~\ref{eq:one-piecewdh} as $F(J,h,\phi)=\widetilde{F}(J, h,\phi,0,0)$, shown as the first quantum circuit in Fig.~\ref{fig:dhonly}(a).
For $h=0$, the expansion of Eq.~\eqref{eq:one-piecewdh} to first order in $\delta h$ only contains terms proportional to $\XI$, $\XZ$, $\IX$ and $\ZX$, as is clear from the Methods section. For this particular case, the explicit form of these error terms is simple enough to be obtained analytically. We write the first order error terms as ($\zeta=A,B$)
\begin{align}
&\frac{\partial\widetilde{F}(J, 0,\phi,\delta h^A,\delta h^B)}{\partial\delta h^\zeta}\nn\\=&\ f_{x0}^\zeta\XI+f_{xz}^\zeta\XZ+f_{0x}^\zeta\IX+f_{zx}^\zeta\ZX,
\label{eq:one-pieceerror}
\end{align}
and for a given $\phi$ value, the coefficients on the right hand side of Eq.~\eqref{eq:one-pieceerror} are only functions of $J/J^{AB}$. Since experimentally it is easy to vary $J$, it then makes sense to choose a value which minimizes the error. Therefore we reiterate the problem as a constrained optimization problem: given the cost function ${\cal K}(J)=\sum_{\zeta=A,B}\left(|f_{x0}^\zeta|^2+|f_{xz}^\zeta|^2+|f_{0x}^\zeta|^2+|f_{zx}^\zeta|^2\right)$ and the constraint that $J\ge J^{AB}$ (which is the physical regime where experiments operate\cite{Shulman.12}), find the value of $J$ which minimizes ${\cal K}(J)$. To simplify expressions of cost functions to be discussed later, we define the ``norm'' of a $4\times4$ matrix $Q$ projected onto the 16-dimensional two-qubit Pauli basis as
$\|Q\|=\left[\sum_{\mu,\nu}\left|{\rm Tr}\ Q(\sigma_\mu\otimes\sigma_\nu)/4\right|^2\right]^{1/2}$ and $\mu,\nu$ run over $\{0,x,y,z\}$ while $\sigma_0=I$. Then
\begin{align}
{\cal K}(J)=\sum_{\zeta=A,B}\|\partial\widetilde{F}/\partial\delta h^\zeta\|^2.\label{eq:Konepiece}
\end{align}
 Here we use the norm of the ``error vector'' in place of the infidelity because the latter requires input of the noise amplitude which we assume to be an unknown constant for a given run.

We show ${\cal K}(J)$ in Fig.~\ref{fig:dhonly}(b). ${\cal K}(J)$ has multiple minima, and one can choose the minimum which is most practical for a specific experiment. Since the curve does not have sharp dips at the minima, even if one is a little off from the optimal value one should still expect error reduction. Results for $J=9.2901J^{AB}$ are shown in  Fig.~\ref{fig:dhonly}(c) as the the red dashed line, which is compared to an unoptimized (arbitrarily chosen) value of $J=3J^{AB}$ for the sequence of Eq.~\eqref{eq:one-piecewdh}. [All curves shown in Fig.~\ref{fig:dhonly}(c) correspond to $\phi=3\pi/4$ and are equivalent to CNOT, and for presentational purposes we have set $\delta h^A=\delta h^B=\delta h$.] Because there is only one instance of the evolution operator $U$ appearing on each side of the $x$-rotations, we term the sequence as ``level-1''. We see that if one simply changes $J$ to the value $9.2901J^{AB}$, the sequence already offers an additional error reduction by two orders of magnitude. Even if we cannot cancel the first-order $\delta h$ error entirely, as is evident from the equal slopes of the lines in Fig.~\ref{fig:dhonly}(c), this reduction in error is already substantial, and it only requires a simple alteration in existing experiments (just changing the $J$ value). Here we also note that if one allows asymmetric pulsing (i.e. $J^A\neq J^B$), the cost function is instead simply ${\cal K}(J^A,J^B)=\left[{\cal K}(J^A)+{\cal K}(J^B)\right]/2$ where ${\cal K}(J)$ is as defined in Eq.~\eqref{eq:Konepiece}. The minima are therefore the same as those for Eq.~\eqref{eq:Konepiece}. We also note that while $J^{AB}$ is assumed to be held constant in producing the figure, our method can easily be adapted to more general situations that can include nontrivial relations between $J^A$, $J^B$ and $J^{AB}$.

Further error reduction may be achieved by adding more parameters in the optimization scheme. We note that
{\begin{widetext}
\begin{align}
&\prod_{j=n}^1\left[U\left(\{J^A_j,0\},\{J^B_j,0\}, J^{AB}_j,{\phi_j}/(2J^{AB}_j)\right)e^{-i\frac{\pi}{2}(\IX+\XI)}\right]
e^{i\frac{\pi}{2}(\IX+\XI)}\notag\\
&\qquad\prod_{j=1}^n
\left[e^{-i\frac{\pi}{2}(\IX+\XI)}U\left(\{J^A_j,0\},\{J^B_j,0\}, J^{AB}_j,{\phi_j}/(2J^{AB}_j)\right)\right]\notag\\
&=\begin{pmatrix}
0 & 0 & 0 & -e^{-i\sum_j\phi_j} \\
0 & 0 &  -e^{i\sum_j\phi_j}  & 0 \\
0 &  -e^{i\sum_j\phi_j}  & 0 & 0 \\
 -e^{-i\sum_j\phi_j}  & 0 & 0 & 0 \\
\end{pmatrix}.\label{eq:multi-piece}
\end{align}
\end{widetext}}
Under our assumption of symmetric dots $J^A_j=J^B_j=J_j$, we have two parameters $J_1$ and $J_2$ for $n=2$, which we call a ``level-2'' sequence. The circuit for this case is shown in Fig.~\ref{fig:dhonly}(a). Defining
\begin{align}
&\widetilde{F}^{(2)}(J_1, J_2, \phi_1,\phi_2,\delta h_1^A,\delta h_1^B,\delta h_2^A,\delta h_2^B)\nn\\
&=U(\{J_2,\delta h_2^A\},\{J_2,\delta h_2^B\}, J^{AB}_2,\phi_2/J^{AB}_2)e^{-i\frac{\pi}{2}(\IX+\XI)}\notag\\
&\times U(\{J_1,\delta h_1^A\},\{J_1,\delta h_1^B\}, J^{AB}_1,\phi_1/J^{AB}_1)e^{-i\frac{\pi}{2}(\IX+\XI)}\notag\\
&\times U(\{J_1,\delta h_1^A\},\{J_1,\delta h_1^B\}, J^{AB}_1,\phi_1/J^{AB}_1)e^{-i\frac{\pi}{2}(\IX+\XI)}\notag\\
&\times U(\{J_2,\delta h_2^A\},\{J_2,\delta h_2^B\}, J^{AB}_2,\phi_2/J^{AB}_2),
\label{eq:two-piecewdh}
\end{align}
the problem now is to minimize the cost function
\begin{align}
{\cal K}(J_1, J_2, \phi_1,\phi_2)=\sum_{\substack{\zeta=A,B\\ j=1,2}}\left\|\frac{\partial \widetilde{F}^{(2)}}{\partial\delta h_j^\zeta}\right\|^2,
\end{align}
subject to the constraint that $\phi_1+\phi_2=N\pi/4$, $\phi_1,\phi_2\ge0$ and $J_1,J_2\ge J^{AB}$. Numerical results for $N=3$ are shown in Fig.~\ref{fig:dhonly}(c) (as in the previous case we have set $\delta h_1^A=\delta h_1^B=\delta h_2^A=\delta h_2^B=\delta h$), where we see that for a wide range of $\delta h$, one may further reduce the error by two orders of magnitude relative to the optimized level-1 sequence. Therefore the pulse sequence is very effective in reducing the error at the cost of extending the sequence by only a factor of two.

\begin{figure*}[t]
\centering\includegraphics[width=1.2\columnwidth]{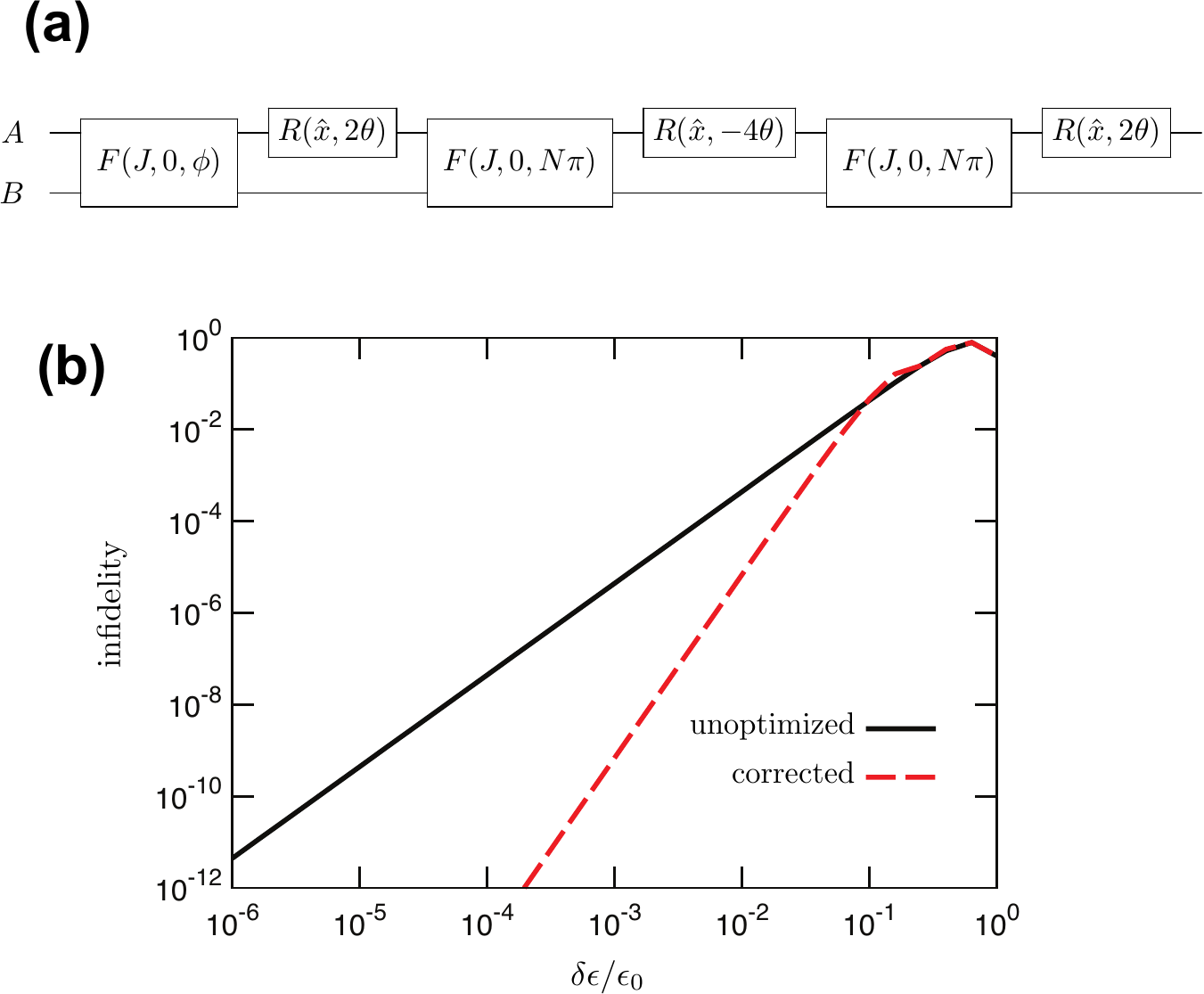}
\caption{Charge noise reduction for $h=0$ while the nuclear noise is absent. (a) Quantum circuit showing the pulse sequence used to cancel the charge noise. (b) Infidelity v.s. charge noise $\delta\epsilon/\epsilon_0$ for unoptimized (black line) and corrected (red dashed line) sequences. For the
unoptimized pulse, we choose $J=3J^{AB}$, $\phi=3\pi/4$, while for the corrected pulse we have  $J=3J^{AB}$, $\phi=3\pi/4$, $N=4$, $\theta=\frac{1}{2}\arccos\left(-\frac{\phi}{2N\pi}\right)$.}
\label{fig:dJonly}
\end{figure*}

\subsection{Charge noise cancellation for $h=0$ case}

As mentioned above, we assume that the charge noise shifts the inter-qubit coupling by $\delta J^{AB}=J^{AB}\delta\epsilon^{AB}/\epsilon_0$. For the $h=0$ case, it does not matter how $J^A$ or $J^B$ responds to the charge noise in the sequence of Eq.~\eqref{eq:one-piece} so long as these effects remain identical before and after the single-qubit $x$-rotations on the left-hand side of  Eq.~\eqref{eq:one-piece}, since $J^A$ and $J^B$ do not appear on the right-hand side. [This also partially explains why Eq.~\eqref{eq:one-piecewdh} produces gates with higher fidelity than a na\"ive implementation of Eq.~\eqref{eq:twoqHam}.] We therefore denote $\delta\epsilon^{AB}/\epsilon_0\equiv\delta\epsilon$ and rewrite the right-hand side of Eq.~\eqref{eq:one-piece} under the influence of charge noise as
\begin{align}
&\widetilde{E}(\phi,\delta\epsilon)\nn\\
=&\ \begin{pmatrix}
0 & 0 & 0 & -e^{-i(1+\delta\epsilon)\phi} \\
0 & 0 &  -e^{i(1+\delta\epsilon)\phi}  & 0 \\
0 &  -e^{i(1+\delta\epsilon)\phi}  & 0 & 0 \\
 -e^{-i(1+\delta\epsilon)\phi}  & 0 & 0 & 0 \\
\end{pmatrix}.
\end{align}

This type of amplitude error also arises in the course of generating Ising gates on two exchange-coupled singlet-triplet qubits.\cite{Kestner.13, Wang.14a} We may therefore adopt an approach similar to what was used in that context and consider a generalization of the SK1 composite pulse sequence:\cite{Brown.04}
\begin{align}
&e^{-i\XI\theta}\widetilde{E}(N\pi,\delta\epsilon)e^{2i\XI\theta}\widetilde{E}(N\pi,\delta\epsilon)e^{-i\XI\theta}\widetilde{E}(\phi,\delta\epsilon)\notag\\
&=\widetilde{E}(\phi,0)\left[\II-i (2N\pi\cos (2 {\theta})+\phi )(\ZZ)
\delta\epsilon\right]\nn\\&\qquad+{\cal O}(\delta\epsilon^2).\label{eq:djonlysk1}
\end{align}
If we choose $\theta=\pm\arccos\left(-\phi/2N\pi\right)/2$, then the charge noise can be completely suppressed to the first order in $\delta\epsilon$. This sequence is shown as the quantum circuit in Fig.~\ref{fig:dJonly}(a). An example of the gate infidelity as a function of the magnitude of charge noise is shown in Fig.~\ref{fig:dJonly}. The difference in slopes of the two curves signifies full cancelation of the first-order error, and one can see that for charge noise at the few-percent level, the error can be reduced by more than an order of magnitude.

\begin{figure*}[t]
\centering\includegraphics[width=1.0\columnwidth]{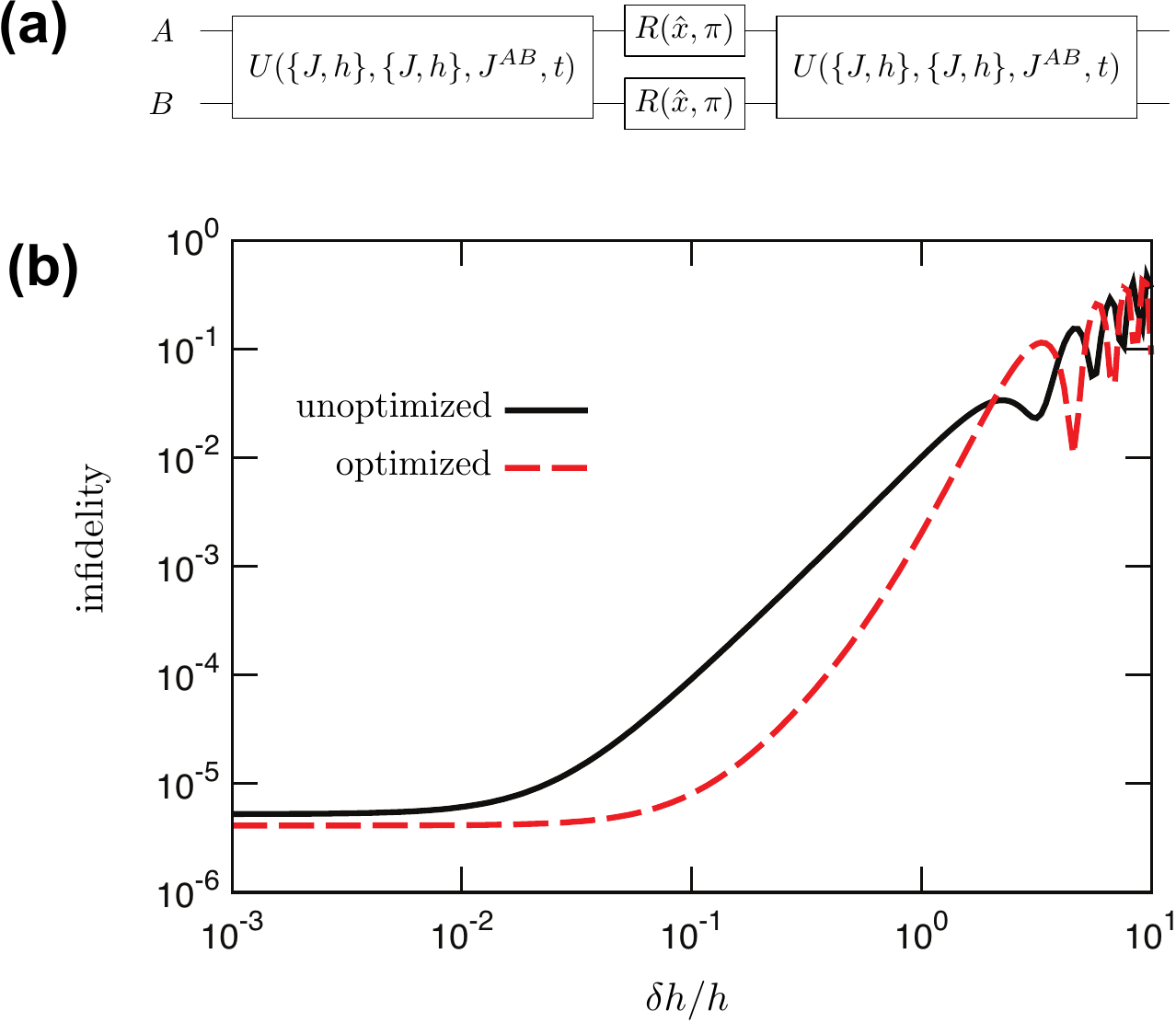}
\caption{Nuclear noise reduction for $h=5J^{AB}$. The charge noise has been set to zero. (a) Quantum circuit showing the pulse sequence used to reduce the nuclear noise. (b) Infidelity v.s. nuclear noise $\delta h/h$ for unoptimized (black line) and optimized (red dashed line) sequences.
Parameters:
Unoptimized pulse: $J=124.83J^{AB}$, $\phi=2J^{AB}t=2.3601$. Optimized pulse: $J=134.20J^{AB}$, $\phi=2.3596$. Both sequences are converted by (noiseless) local operations to CNOT to calculate the infidelity.}
\label{fig:dhfiniteh}
\end{figure*}

\subsection{Reducing nuclear noise for nonzero $h$}

In actual experiments, one needs a nonzero $h$ to generate $x$-rotations, so we must depart from the ideal $h=0$ limit discussed above. In this case, the right-hand side of Eq.~\eqref{eq:one-piece} is no longer an anti-diagonal matrix. Since its explicit analytical form is complicated, one must check its equivalence to a CNOT gate by directly evaluating the Makhlin invariant. Taking the ``level-1'' sequence [Eq.~\eqref{eq:one-piecewdh}] as an example, the cost function now has a given nonzero $h$ value: ${\cal K}(J,\phi)=\|\partial\widetilde{F}(J,h,\phi,\delta h)/\partial\delta h\|^2$, but most importantly, we have an extra constraint in that $\widetilde{F}$ must be equivalent to CNOT. To impose this constraint, we define a function ${\cal M}(Q)=\left|G_1\right|^2+\left|G_2-1\right|^2$, where $Q$ is a $4\times4$ unitary, and $G_1, G_2$ are the Makhlin invariants associated with it. This function may be viewed as the ``distance'' between the net evolution generated by any pulse sequence and the desired entangling gate.

Because $\phi=J^{AB}t$ no longer solely determines whether the gate is equivalent to CNOT, we will use $\phi$ as a search parameter in addition to $J$. Our optimization problem thus becomes the minimization of ${\cal K}(J,\phi)$ subject to the constraints $J\ge J^{AB}$, $\phi>0$ and the additional constraint from the Makhlin invariance  $\left|{\cal M}(\widetilde{F})\right|\le\eta$, with $\eta$ a small dimensionless number which we take to be $10^{-10}$ in our calculations. This additional constraint makes a brute-force numerical optimization very difficult even if there are now more free parameters, because during the search a given point in the parameter search space is only allowed to evolve in a very narrow strip determined by $\eta$. We therefore do not find such a brute force direct strategy to be practical here. Instead, we adopt a more practical strategy in which we use the known result for $h=0$ and slowly turn on $h$, increasing it in small increments of size $\Delta h$ towards a desired value. This practice ensures that with each $h$ value, the search initiates from the solution for $h-\Delta h$,  which is presumably not far away from the optimal solution for $h$, while at the same time satisfying the constraint of Makhlin invariance. This iterative discretized numerical strategy seems to be practical and should give the same answer as a direct numerical brute force technique.

A typical result for $h=5J^{AB}$ is shown in Fig.~\ref{fig:dhfiniteh}. Fig.~\ref{fig:dhfiniteh}(a) shows the quantum circuit used in this case, and Fig.~\ref{fig:dhfiniteh}(b) depicts the dependence of gate infidelity on the nuclear noise level. An order-of-magnitude reduction in error can be seen for a range of $\delta h/h$ around 10\%. In producing the gate infidelity, we convert the output of the sequence of Eq.~\eqref{eq:one-piecewdh} without noise to CNOT via local operations. Namely, we find real vectors $\boldsymbol{r}^A_L$, $\boldsymbol{r}^B_L$, $\boldsymbol{r}^A_R$, $\boldsymbol{r}^B_R$, such that
\begin{align}
&\exp\left\{-i\left[\left(\boldsymbol{r}^A_L\cdot\boldsymbol{\sigma}\right)\otimes I+I\otimes\left(\boldsymbol{r}^B_L\cdot\boldsymbol{\sigma}\right)\right]\right\}\cdot Q\nn\\
&\cdot\exp\left\{-i\left[\left(\boldsymbol{r}^A_R\cdot\boldsymbol{\sigma}\right)\otimes I+I\otimes\left(\boldsymbol{r}^B_R\cdot\boldsymbol{\sigma}\right)\right]\right\}\notag\\&={\rm CNOT}
\end{align}
where $\boldsymbol{\sigma}=\left\{\sigma_x,\sigma_y,\sigma_z\right\}$. (The local operations are assumed to be free from noise.) As a consequence, the extent to which the entangling gate $Q$ resembles CNOT is manifest in Fig.~\ref{fig:dhfiniteh} as a non-zero saturation value for small $\delta h$. This error is on the order of $\sqrt{\eta}$, and as long as $\eta<10^{-8}$, it is small enough to not hinder quantum error correction. We also note that the $J$ values used for both unoptimized and optimized pulses in Fig.~\ref{fig:dhfiniteh} are larger compared to Fig.~\ref{fig:dhonly}. These larger values are required in the nonzero $h$ case to ensure $Q$ is close enough to CNOT, i.e., within the prescribed precision set by $\eta$. Smaller $J$ may certainly be used at the cost of increasing $\eta$, and we have verified that solutions can be found for most cases as long as $\eta<10^{-6}$.

\begin{figure*}[t]
\centering\includegraphics[width=1.9\columnwidth]{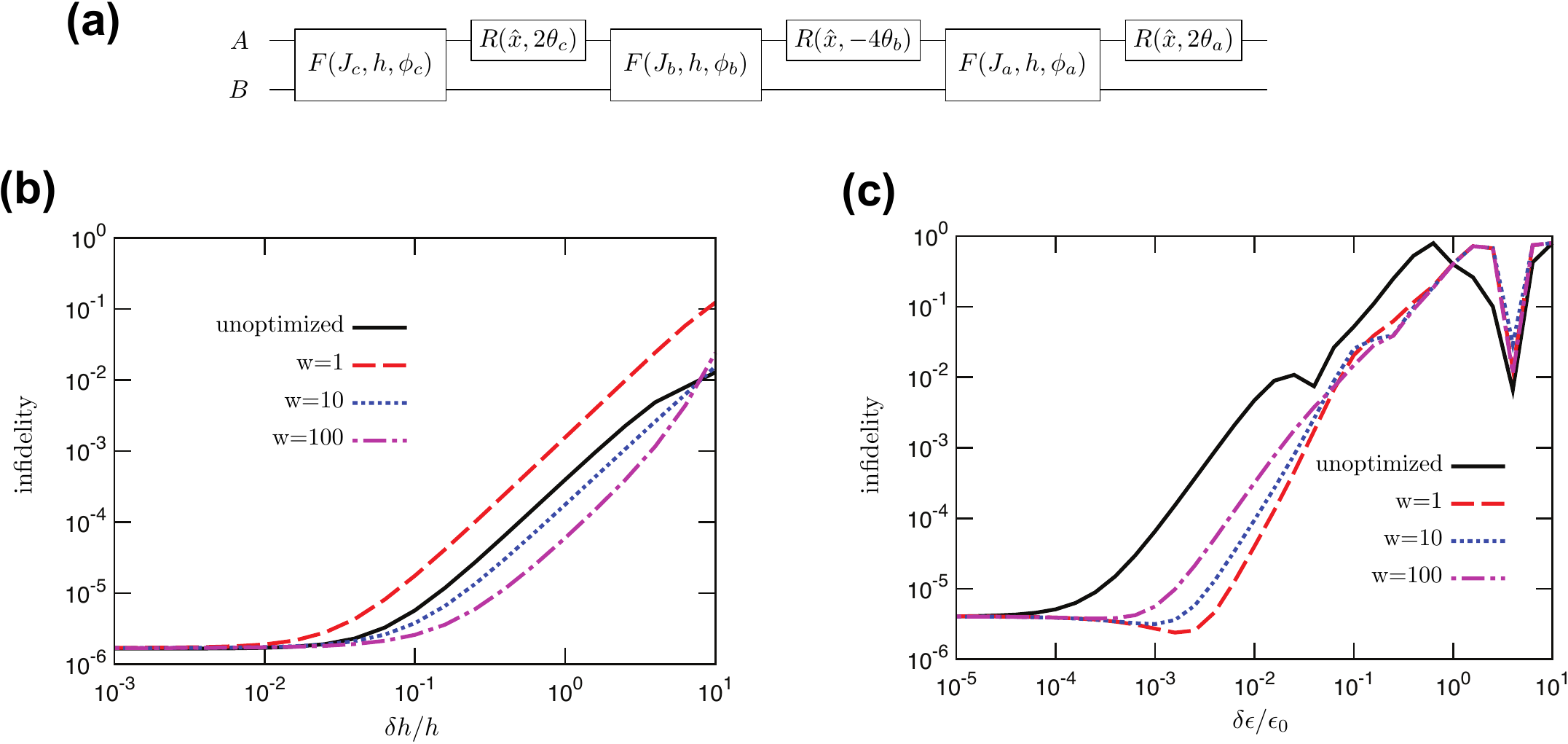}
\caption{Simultaneous reduction of nuclear and charge noise for  $h=2.4J^{AB}$. (a) Quantum circuit showing the pulse sequence used to reduce the noise. (b) Infidelity v.s. nuclear noise $\delta h/h$ with charge noise fixed at zero. (c) Infidelity v.s. charge noise $\delta\epsilon/\epsilon_0$ with zero nuclear noise.
Parameters:
Unoptimized pulse: $J=64.128J^{AB}$, $\phi=2.3596$.
Optimized pulse with $w=1$:
$J_a=151.56J^{AB}$, $J_b=58.609J^{AB}$, $J_c=33.712J^{AB}$, $\phi_a=0.21612$, $\phi_b=0.35106$, $\phi_c=0.22303$, $\theta_a=14.200$, $\theta_b=1.5999$ and $\theta_c=3.0629$.
Optimized pulse with $w=10$:
$J_a=89.590J^{AB}$, $J_b=83.522J^{AB}$, $J_c=38.413J^{AB}$, $\phi_a=0.33514$, $\phi_b=0.27385$, $\phi_c=0.17992$, $\theta_a=22.690$, $\theta_b=1.5253$ and $\theta_c=3.0535$.
Optimized pulse with $w=100$:
$J_a=115.02J^{AB}$, $J_b=64.424J^{AB}$, $J_c=36.711J^{AB}$, $\phi_a=0.24015$, $\phi_b=0.36612$, $\phi_c=0.18231$, $\theta_a=12.940$, $\theta_b=1.5308$ and $\theta_c=3.1414$.
}
\label{fig:dhdJfiniteh}
\end{figure*}

\subsection{Simultaneous reduction of nuclear noise and charge noise for nonzero $h$} To make the sequence as robust as possible against both nuclear noise and charge noise, we first rewrite the longer sequence of Eq.~\eqref{eq:djonlysk1} as
\begin{align}
\widetilde{D}&=e^{-2i\XI\theta_a}F(J_a, h,\phi_a)e^{4i\XI\theta_b}F(J_b,h,\phi_b)\nn\\
&\quad\times e^{-2i\XI\theta_c}F(J_c, h,\phi_c)\label{eq:simulcompseq}
\end{align}
and we now have nine free parameters that can be tuned during the optimization: $J_a$, $J_b$, $J_c$, $\phi_a$, $\phi_b$, $\phi_c$, $\theta_a$, $\theta_b$, and $\theta_c$.  Again, our strategy is to start from the known solution at $h=0$ and then slowly turn on $h$ to ensure that the search always starts from a near-optimal solution while satisfying the constraint  $\left|{\cal M}(\widetilde{D})\right|\le\eta$ as much as possible. In addition to the Makhlin invariant, other constraints include $J_{a,b,c}\ge J^{AB}$, $\phi_{a,b,c}\ge0$. There are no constraints on $\theta_{a,b,c}$ since they denote exact single-qubit operations.

The cost function contains contributions from both the nuclear noise and the charge noise, but since they are not directly comparable, one must assign a weight $w$ when summing the two contributions in the cost function. We may therefore write the cost function as
\begin{align}
{\cal K}=\left\|w\frac{\partial\widetilde{D}}{\partial\delta h}\right\|^2+\left\|\frac{\partial\widetilde{D}}{\partial\delta\epsilon}\right\|^2.
\end{align}
In practice, $w$ should be chosen according to whether charge noise or nuclear noise is responsible for most of the decoherence. Increasing $w$ means that one is willing to sacrifice a portion of the charge noise reduction in exchange for an additional reduction in nuclear noise, and vice versa. The results for several $w$ values are shown in Fig.~\ref{fig:dhdJfiniteh}. Fig.~\ref{fig:dhdJfiniteh}(a) shows the quantum circuit used.  Fig.~\ref{fig:dhdJfiniteh}(b) shows the dependence of the infidelity on nuclear noise for zero charge noise, while Fig.~\ref{fig:dhdJfiniteh}(c) shows how the infidelity changes in accordance with the charge noise when the nuclear noise is absent. As in the previous section, we convert the resulting gate in the absence of noise to CNOT before evaluating the infidelities, so that the error resulting from the Makhlin constraint characterized by $\eta$ may be seen as part of the infidelities. For $w=1$, reduction of the charge noise is most pronounced, but the sequence offers no reduction in nuclear noise. The error actually increases from the unoptimized short sequence, due to the complexity of Eq.~\eqref{eq:simulcompseq} compared to Eq.~\eqref{eq:one-piecewdh}. As expected, when one increases $w$, the sequence starts to cancel nuclear noise, at the cost of offering less (but still acceptable) reduction in charge noise. For $w=100$, the sequence reduces both nuclear noise and charge noise by approximately one order of magnitude. The added flexibility afforded by $w$ is not only beneficial in experiments where one type of noise dominates over the other, but also in situations where one wishes to simultaneously employ alternative methods for suppressing noise, such as Bayesian estimation {\cite{Shulman.14}}.

\section{Discussion}

The method presented here can be easily extended to generate entangling gates other than CNOT or applied to systems other than the singlet-triplet qubits discussed here. If one simply wishes to generate an arbitrary entangling gate in a noise-resistant fashion, then the constraint imposed by the Makhlin invariants can be removed, and it should be much easier to find optimal driving parameters. (The invariants must still be monitored to ensure that a non-entangling product of local gates is not generated.) Very recent experiments in double-quantum-dot charge qubits have demonstrated a CNOT gate with a fidelity comparable to that obtained for singlet-triplet spin qubits,\cite{Li.14} and the coupling Hamiltonian is identical to Eq.~\eqref{eq:twoqHam} when $J^A,J^B$ are interpreted as the energy splittings between charge states and $h^A, h^B$ as the inter-dot tunnelings. In this case, the energy splittings are no longer constrained to be strictly positive, and their dependence on the detuning is typically linear. These differences can easily be incorporated into our optimization procedure, and all the analysis presented here, including the structure of the pulse sequences, remains the same. We further note that our method may be generalized to produce noise-suppressing pulse sequences for other types of qubits, such as the exchange-only qubit or the resonant-exchange qubit,\cite{DiVincenzo.00,Medford.13,Pal.14} as well as the hybrid qubit demonstrated recently,\cite{Shi.12,Kim.14} which can be coupled capacitively.
We also emphasise that several assumptions that we have made for the purposes of presentation, including symmetric pulsing and the exponential charge noise model, can easily be lifted to address more general situations. For example, asymmetric pulsing may be treated by adding search parameters, while for a general charge noise model the sequence of Eq.~\eqref{eq:djonlysk1} still holds as long as the auxiliary angle $\theta$ is determined by the actual amplitude errors of $\widetilde{E}(N\pi,\delta\epsilon)$ and $\widetilde{E}(\phi,\delta\epsilon)$. Discussions for similar cases in exchange-coupled qubits may be found in Ref.~\onlinecite{Wang.14a}. Ref.~{\onlinecite{Wang.14a}} also treats a case where the static-noise approximation is lifted, and has found that as long as the noise is concentrated at low frequencies, dynamically corrected single-qubit gates work fine. In future work, it would be interesting to map out precisely the types of noise spectra for which our two-qubit gates do and do not work well, although it will be computationally much more expensive than in the case of single-qubit gates.

Precise, error-free entangling gate operations on coupled qubits is currently the bottleneck for scalable quantum computation using solid state spin qubits.  Our results show how to reduce both nuclear noise and charge noise while at the same time offering the advantage that the sequences are simple enough for immediate experimental implementation. The simpler sequence for nuclear noise reduction in the case of negligible Overhauser field gradients only requires tuning the value of the exchange coupling in the existing experimental control protocol,\cite{Shulman.12} while our more powerful sequence that suppresses both types of noise extends that sequence merely by a factor of three.  We therefore believe that our results are of immediate practical use and constitute an important step toward scalable semiconductor quantum technologies based on coupled spin qubits.

\onecolumngrid

%\begin{widetext}

\section{Methods}

{\bf{Explicit expression for two-qubit evolution}}
The evolution for a time $t$ under the Hamiltonian, Eq.~\eqref{eq:twoqHam}, may be expressed analytically as:
\allowdisplaybreaks[2]
\begin{eqnarray}
U&=&\sum_{k=1}^4\frac{e^{r_kt}}{4A_k}\bigg(A_kI\otimes I+u_{ix,k} I\otimes\sigma_x+u_{xi,k}\sigma_x\otimes I+u_{iz,k} I\otimes\sigma_z+u_{zi,k}\sigma_z\otimes I\nn\\
&&+u_{xz,k}\sigma_x\otimes\sigma_z+u_{zx,k}\sigma_z\otimes\sigma_x+u_{xx,k}\sigma_x\otimes\sigma_x+ u_{yy,k}\sigma_y\otimes\sigma_y+u_{zz,k}\sigma_z\otimes\sigma_z\bigg),
\end{eqnarray}
where we have defined
\begin{eqnarray}
A_k&=&\left[(h^A)^2+(h^B)^2+(J^A)^2+(J^B)^2+(J^{AB})^2\right] r_k-2 i J^A J^B J^{AB}+r_k^3,\nn\\
u_{ix,k}&=&-i h^B \left[-(h^A)^2+(h^B)^2-(J^A)^2+(J^B)^2+(J^{AB})^2+r_k^2\right],\nn\\
u_{xi,k}&=&u_{ix,k}(J^A\leftrightarrow J^B,h^A\leftrightarrow h^B),\nn\\
u_{iz,k}&=&-i \left\{-J^B \left[(h^A)^2-(h^B)^2+(J^A)^2+(J^{AB})^2\right]+J^B r_k^2-2 i J^A J^{AB} r_k+(J^B)^3\right\},\nn\\
u_{zi,k}&=&u_{iz,k}(J^A\leftrightarrow J^B,h^A\leftrightarrow h^B),\nn\\
u_{xz,k}&=&2 i h^A \left(J^A J^{AB}+i J^B r_k\right),\nn\\
u_{zx,k}&=&u_{xz,k}(J^A\leftrightarrow J^B,h^A\leftrightarrow h^B),\nn\\
u_{xx,k}&=&-2 h^A h^B r_k,\nn\\
u_{yy,k}&=&-2 i h^A h^B J^{AB},\nn\\
u_{zz,k}&=&-i \left\{J^{AB} \left[(h^A)^2+(h^B)^2-(J^A)^2-(J^B)^2+(J^{AB})^2\right]+J^{AB} r_k^2-2 i J^A J^B r_k\right\}.
\end{eqnarray}
\twocolumngrid
The $r_k$ are the roots of the following polynomial:
\begin{equation}
P(r)=a+br+cr^2+r^4,
\end{equation}

\allowdisplaybreaks[1]

with
\begin{align}
a&=2 (J^{AB})^2 \left[(h^A)^2+(h^B)^2-(J^A)^2-(J^B)^2\right]\nn\\
&\qquad+\left[(h^A)^2-(h^B)^2+(J^A)^2-(J^B)^2\right]^2+(J^{AB})^4,\nn\\
b&=-8 i J^A J^B J^{AB},\nn\\
c&=2 \left[(h^A)^2+(h^B)^2+(J^A)^2+(J^B)^2+(J^{AB})^2\right].
\end{align}

\section*{Acknowledgement} This work is supported by LPS-CMTC and IARPA-MQCO grants.


\begin{thebibliography}{10}

\expandafter\ifx\csname url\endcsname\relax
  \def\url#1{\texttt{#1}}\fi
\expandafter\ifx\csname urlprefix\endcsname\relax\def\urlprefix{URL }\fi
\providecommand{\bibinfo}[2]{#2}
\providecommand{\eprint}[2][]{\url{#2}}


\bibitem{NielsenChuang.00}
\bibinfo{author}{Nielsen, M.} \& \bibinfo{author}{Chuang, I.}
\newblock \emph{\bibinfo{title}{Quantum Computation and Quantum Information}}
  (\bibinfo{publisher}{Cambridge University Press, Cambridge},
  \bibinfo{year}{2000}).

\bibitem{DiCarlo.09}
\bibinfo{author}{DiCarlo, L.} \emph{et~al.}
\newblock \bibinfo{title}{Demonstration of two-qubit algorithms with a
  superconducting quantum processor}.
\newblock \emph{\bibinfo{journal}{Nature}} \textbf{\bibinfo{volume}{460}},
  \bibinfo{pages}{240--244} (\bibinfo{year}{2009}).

\bibitem{Gulde.03}
\bibinfo{author}{Gulde, S.} \emph{et~al.}
\newblock \bibinfo{title}{Implementation of the {Deutsch-Jozsa} algorithm on an
  ion-trap quantum computer}.
\newblock \emph{\bibinfo{journal}{Nature}} \textbf{\bibinfo{volume}{421}},
  \bibinfo{pages}{48--50} (\bibinfo{year}{2003}).

\bibitem{O'Brien.03}
\bibinfo{author}{O'Brien, J.~L.}, \bibinfo{author}{Pryde, G.~J.},
  \bibinfo{author}{White, A.~G.}, \bibinfo{author}{Ralph, T.~C.} \&
  \bibinfo{author}{Branning, D.}
\newblock \bibinfo{title}{Demonstration of an all-optical quantum
  controlled-not gate}.
\newblock \emph{\bibinfo{journal}{Nature}} \textbf{\bibinfo{volume}{426}},
  \bibinfo{pages}{264--267} (\bibinfo{year}{2003}).

\bibitem{Nowack.11}
\bibinfo{author}{Nowack, K.~C.} \emph{et~al.}
\newblock \bibinfo{title}{Single-shot correlations and two-qubit gate of
  solid-state spins}.
\newblock \emph{\bibinfo{journal}{Science}} \textbf{\bibinfo{volume}{333}},
  \bibinfo{pages}{1269--1272} (\bibinfo{year}{2011}).

\bibitem{Brunner.11}
\bibinfo{author}{Brunner, R.} \emph{et~al.}
\newblock \bibinfo{title}{Two-qubit gate of combined single-spin rotation and
  interdot spin exchange in a double quantum dot}.
\newblock \emph{\bibinfo{journal}{Phys. Rev. Lett.}}
  \textbf{\bibinfo{volume}{107}}, \bibinfo{pages}{146801}
  (\bibinfo{year}{2011}).

\bibitem{Veldhorst.14}
\bibinfo{author}{{Veldhorst}, M.} \emph{et~al.}
\newblock \bibinfo{title}{A two qubit logic gate in silicon.}
  \bibinfo{pages}{Preprint at $\langle$http://arxiv.org/abs/1411.5760$\rangle$}
  (\bibinfo{year}{2014}).

\bibitem{Kim.14}
\bibinfo{author}{{Kim}, D.} \emph{et~al.}
\newblock \bibinfo{title}{{Quantum control and process tomography of a
  semiconductor quantum dot hybrid qubit}}.
\newblock \emph{\bibinfo{journal}{Nature}} \textbf{\bibinfo{volume}{511}},
  \bibinfo{pages}{70--74} (\bibinfo{year}{2014}).

\bibitem{Muhonen.14}
\bibinfo{author}{Muhonen, J.~T.} \emph{et~al.}
\newblock \bibinfo{title}{Storing quantum information for 30 seconds in a
  nanoelectronic device}.
\newblock \emph{\bibinfo{journal}{Nat. Nanotechnol.}}
  \textbf{\bibinfo{volume}{9}}, \bibinfo{pages}{986--991}
  (\bibinfo{year}{2014}).

\bibitem{Taylor.05}
\bibinfo{author}{Taylor, J.~M.} \emph{et~al.}
\newblock \bibinfo{title}{Fault-tolerant architecture for quantum computation
  using electrically controlled semiconductor spins}.
\newblock \emph{\bibinfo{journal}{Nature Phys.}} \textbf{\bibinfo{volume}{1}},
  \bibinfo{pages}{177--183} (\bibinfo{year}{2005}).

\bibitem{Petta.05}
\bibinfo{author}{Petta, J.} \emph{et~al.}
\newblock \bibinfo{title}{{Coherent manipulation of coupled electron spins in
  semiconductor quantum dots}}.
\newblock \emph{\bibinfo{journal}{{Science}}} \textbf{\bibinfo{volume}{{309}}},
  \bibinfo{pages}{{2180--2184}} (\bibinfo{year}{{2005}}).

\bibitem{Foletti.09}
\bibinfo{author}{Foletti, S.}, \bibinfo{author}{Bluhm, H.},
  \bibinfo{author}{Mahalu, D.}, \bibinfo{author}{Umansky, V.} \&
  \bibinfo{author}{Yacoby, A.}
\newblock \bibinfo{title}{{Universal quantum control of two-electron spin
  quantum bits using dynamic nuclear polarization}}.
\newblock \emph{\bibinfo{journal}{{Nature Phys.}}}
  \textbf{\bibinfo{volume}{{5}}}, \bibinfo{pages}{{903--908}}
  (\bibinfo{year}{{2009}}).

\bibitem{Bluhm.11}
\bibinfo{author}{Bluhm, H.} \emph{et~al.}
\newblock \bibinfo{title}{{Dephasing time of GaAs electron-spin qubits coupled
  to a nuclear bath exceeding 200 $\mu s$}}.
\newblock \emph{\bibinfo{journal}{{Nature Phys.}}}
  \textbf{\bibinfo{volume}{{7}}}, \bibinfo{pages}{{109--113}}
  (\bibinfo{year}{{2011}}).

\bibitem{Shulman.14}
\bibinfo{author}{{Shulman}, M.~D.} \emph{et~al.}
\newblock \bibinfo{title}{{Suppressing qubit dephasing using real-time
  Hamiltonian estimation}}.
\newblock \emph{\bibinfo{journal}{Nat. Commun.}} \textbf{\bibinfo{volume}{5}},
  \bibinfo{pages}{5156} (\bibinfo{year}{2014}).

\bibitem{Klinovaja.12}
\bibinfo{author}{Klinovaja, J.}, \bibinfo{author}{Stepanenko, D.},
  \bibinfo{author}{Halperin, B.} \& \bibinfo{author}{Loss, D.}
\newblock \bibinfo{title}{Exchange-based CNOT gates for singlet-triplet qubits
  with spin-orbit interaction}.
\newblock \emph{\bibinfo{journal}{Phys. Rev. B}} \textbf{\bibinfo{volume}{86}},
  \bibinfo{pages}{085423} (\bibinfo{year}{2012}).

\bibitem{Kestner.13}
\bibinfo{author}{Kestner, J.~P.}, \bibinfo{author}{Wang, X.},
  \bibinfo{author}{Bishop, L.~S.}, \bibinfo{author}{Barnes, E.} \&
  \bibinfo{author}{Das~Sarma, S.}
\newblock \bibinfo{title}{Noise-resistant control for a spin qubit array}.
\newblock \emph{\bibinfo{journal}{Phys. Rev. Lett.}}
  \textbf{\bibinfo{volume}{110}}, \bibinfo{pages}{140502}
  (\bibinfo{year}{2013}).

\bibitem{vanWeperen.11}
\bibinfo{author}{van Weperen, I.} \emph{et~al.}
\newblock \bibinfo{title}{Charge-state conditional operation of a spin qubit}.
\newblock \emph{\bibinfo{journal}{Phys. Rev. Lett.}}
  \textbf{\bibinfo{volume}{107}}, \bibinfo{pages}{030506}
  (\bibinfo{year}{2011}).

\bibitem{Trifunovic.12}
\bibinfo{author}{Trifunovic, L.} \emph{et~al.}
\newblock \bibinfo{title}{Long-distance spin-spin coupling via floating gates}.
\newblock \emph{\bibinfo{journal}{Phys. Rev. X}} \textbf{\bibinfo{volume}{2}},
  \bibinfo{pages}{011006} (\bibinfo{year}{2012}).

\bibitem{Shulman.12}
\bibinfo{author}{Shulman, M.~D.} \emph{et~al.}
\newblock \bibinfo{title}{Demonstration of entanglement of electrostatically
  coupled singlet-triplet qubits}.
\newblock \emph{\bibinfo{journal}{Science}} \textbf{\bibinfo{volume}{336}},
  \bibinfo{pages}{202--205} (\bibinfo{year}{2012}).

\bibitem{Wang.12}
\bibinfo{author}{{Wang}, X.} \emph{et~al.}
\newblock \bibinfo{title}{{Composite pulses for robust universal control of
  singlet-triplet qubits}}.
\newblock \emph{\bibinfo{journal}{Nat. Commun.}} \textbf{\bibinfo{volume}{3}},
  \bibinfo{pages}{997} (\bibinfo{year}{2012}).

\bibitem{Grace.12}
\bibinfo{author}{Grace, M.}, \bibinfo{author}{Dominy, J.},
  \bibinfo{author}{Witzel, W.} \& \bibinfo{author}{Carroll, M.}
\newblock \bibinfo{title}{Optimized pulses for the control of uncertain
  qubits}.
\newblock \emph{\bibinfo{journal}{Phys. Rev. A}} \textbf{\bibinfo{volume}{85}},
  \bibinfo{pages}{052313} (\bibinfo{year}{2012}).

\bibitem{Wang.14a}
\bibinfo{author}{Wang, X.}, \bibinfo{author}{Bishop, L.~S.},
  \bibinfo{author}{Barnes, E.}, \bibinfo{author}{Kestner, J.~P.} \&
  \bibinfo{author}{Sarma, S.~D.}
\newblock \bibinfo{title}{Robust quantum gates for singlet-triplet spin qubits
  using composite pulses}.
\newblock \emph{\bibinfo{journal}{Phys. Rev. A}} \textbf{\bibinfo{volume}{89}},
  \bibinfo{pages}{022310} (\bibinfo{year}{2014}).

\bibitem{Barnes.14}
\bibinfo{author}{{Barnes}, E.}, \bibinfo{author}{{Wang}, X.} \&
  \bibinfo{author}{{Das Sarma}, S.}
\newblock \bibinfo{title}{{Robust quantum control using smooth pulses and
  topological winding.}} \bibinfo{pages}{Preprint at
  $\langle$http://arxiv.org/abs/1409.7063$\rangle$} (\bibinfo{year}{2014}).

\bibitem{Jones.03}
\bibinfo{author}{Jones, J.~A.}
\newblock \bibinfo{title}{Robust ising gates for practical quantum
  computation}.
\newblock \emph{\bibinfo{journal}{Phys. Rev. A}} \textbf{\bibinfo{volume}{67}},
  \bibinfo{pages}{012317} (\bibinfo{year}{2003}).

\bibitem{Calderon.14}
\bibinfo{author}{Calderon-Vargas, F.~A.} \& \bibinfo{author}{Kestner, J.~P.}
\newblock \bibinfo{title}{Directly accessible entangling gates for capacitively
  coupled singlet-triplet qubits}.
\newblock \emph{\bibinfo{journal}{Phys. Rev. B}} \textbf{\bibinfo{volume}{91}},
  \bibinfo{pages}{035301} (\bibinfo{year}{2015}).

\bibitem{Srinivasa.14}
\bibinfo{author}{{Srinivasa}, V.} \& \bibinfo{author}{{Taylor}, J.~M.}
\newblock \bibinfo{title}{{Capacitively coupled singlet-triplet qubits in the
  double charge resonant regime.}} \bibinfo{pages}{Preprint at
  $\langle$http://arxiv.org/abs/1408.4740$\rangle$} (\bibinfo{year}{2014}).

\bibitem{Johnson.05}
\bibinfo{author}{Johnson, A.~C.} \emph{et~al.}
\newblock \bibinfo{title}{Triplet-singlet spin relaxation via nuclei in a
  double quantum dot}.
\newblock \emph{\bibinfo{journal}{Nature}} \textbf{\bibinfo{volume}{435}},
  \bibinfo{pages}{925--928} (\bibinfo{year}{2005}).

\bibitem{Medford.12}
\bibinfo{author}{Medford, J.} \emph{et~al.}
\newblock \bibinfo{title}{Scaling of dynamical decoupling for spin qubits}.
\newblock \emph{\bibinfo{journal}{Phys. Rev. Lett.}}
  \textbf{\bibinfo{volume}{108}}, \bibinfo{pages}{086802}
  (\bibinfo{year}{2012}).

\bibitem{Reilly.08}
\bibinfo{author}{Reilly, D.~J.} \emph{et~al.}
\newblock \bibinfo{title}{{Measurement of temporal correlations of the
  Overhauser field in a double quantum dot}}.
\newblock \emph{\bibinfo{journal}{Phys. Rev. Lett.}}
  \textbf{\bibinfo{volume}{101}}, \bibinfo{pages}{236803}
  (\bibinfo{year}{2008}).

\bibitem{Dial.13}
\bibinfo{author}{Dial, O.} \emph{et~al.}
\newblock \bibinfo{title}{Charge noise spectroscopy using coherent exchange
  oscillations in a singlet-triplet qubit}.
\newblock \emph{\bibinfo{journal}{Phys. Rev. Lett.}}
  \textbf{\bibinfo{volume}{110}}, \bibinfo{pages}{146804}
  (\bibinfo{year}{2013}).

\bibitem{Maune.12}
\bibinfo{author}{Maune, B.~M.} \emph{et~al.}
\newblock \bibinfo{title}{{Coherent singlet-triplet oscillations in a
  silicon-based double quantum dot}}.
\newblock \emph{\bibinfo{journal}{{Nature}}} \textbf{\bibinfo{volume}{{481}}},
  \bibinfo{pages}{{344--347}} (\bibinfo{year}{{2012}}).

\bibitem{Wu.14}
\bibinfo{author}{Wu, X.} \emph{et~al.}
\newblock \bibinfo{title}{Two-axis control of a singlet-triplet qubit with an
  integrated micromagnet}.
\newblock \emph{\bibinfo{journal}{Proc. Natl. Acad. Sci. USA}}
  \textbf{\bibinfo{volume}{111}}, \bibinfo{pages}{11938--11942}
  (\bibinfo{year}{2014}).

\bibitem{Wang.14b}
\bibinfo{author}{Wang, X.} \emph{et~al.}
\newblock \bibinfo{title}{Noise-compensating pulses for electrostatically
  controlled silicon spin qubits}.
\newblock \emph{\bibinfo{journal}{Phys. Rev. B}} \textbf{\bibinfo{volume}{90}},
  \bibinfo{pages}{155306} (\bibinfo{year}{2014}).

\bibitem{Makhlin.02}
\bibinfo{author}{Makhlin, Y.}
\newblock \bibinfo{title}{Nonlocal properties of two-qubit gates and mixed
  states, and the optimization of quantum computations}.
\newblock \emph{\bibinfo{journal}{Quantum Inf. Process.}}
  \textbf{\bibinfo{volume}{1}}, \bibinfo{pages}{243--252}
  (\bibinfo{year}{2002}).

\bibitem{Brown.04}
\bibinfo{author}{Brown, K.}, \bibinfo{author}{Harrow, A.} \&
  \bibinfo{author}{Chuang, I.}
\newblock \bibinfo{title}{Arbitrarily accurate composite pulse sequences}.
\newblock \emph{\bibinfo{journal}{Phys. Rev. A}} \textbf{\bibinfo{volume}{70}},
  \bibinfo{pages}{052318} (\bibinfo{year}{2004}).

\bibitem{Li.14}
\bibinfo{author}{{Li}, H.-O.} \emph{et~al.}
\newblock \bibinfo{title}{{Controlled-NOT quantum logic gate in two strongly
  coupled semiconductor charge qubits.}} \bibinfo{pages}{Preprint at
  $\langle$http://arxiv.org/abs/1411.2177$\rangle$} (\bibinfo{year}{2014}).

\bibitem{DiVincenzo.00}
\bibinfo{author}{DiVincenzo, D.~P.}, \bibinfo{author}{Bacon, D.},
  \bibinfo{author}{Kempe, J.}, \bibinfo{author}{Burkard, G.} \&
  \bibinfo{author}{Whaley, K.~B.}
\newblock \bibinfo{title}{Universal quantum computation with the exchange
  interaction}.
\newblock \emph{\bibinfo{journal}{Nature}} \textbf{\bibinfo{volume}{408}},
  \bibinfo{pages}{339--342} (\bibinfo{year}{2000}).

\bibitem{Medford.13}
\bibinfo{author}{Medford, J.} \emph{et~al.}
\newblock \bibinfo{title}{Quantum-dot-based resonant exchange qubit}.
\newblock \emph{\bibinfo{journal}{Phys. Rev. Lett.}}
  \textbf{\bibinfo{volume}{111}}, \bibinfo{pages}{050501}
  (\bibinfo{year}{2013}).

\bibitem{Pal.14}
\bibinfo{author}{Pal, A.}, \bibinfo{author}{Rashba, E.~I.} \&
  \bibinfo{author}{Halperin, B.~I.}
\newblock \bibinfo{title}{Driven nonlinear dynamics of two coupled
  exchange-only qubits}.
\newblock \emph{\bibinfo{journal}{Phys. Rev. X}} \textbf{\bibinfo{volume}{4}},
  \bibinfo{pages}{011012} (\bibinfo{year}{2014}).

\bibitem{Shi.12}
\bibinfo{author}{Shi, Z.} \emph{et~al.}
\newblock \bibinfo{title}{Fast hybrid silicon double-quantum-dot qubit}.
\newblock \emph{\bibinfo{journal}{Phys. Rev. Lett.}}
  \textbf{\bibinfo{volume}{108}}, \bibinfo{pages}{140503}
  (\bibinfo{year}{2012}).

\end{thebibliography}
\end{document}